\documentclass[10pt,conference,epsfig]{IEEEtran}
\usepackage{graphicx}
\usepackage{subfigure,color}
\usepackage{amssymb,varwidth}
\usepackage{cite,tcolorbox}
\usepackage{amsmath,setspace}
\usepackage{bm,comment,epstopdf}
\usepackage{algorithm}
\usepackage{algpseudocode}
\makeatletter

\newcommand{\Rmnum}[1]{\expandafter\@slowromancap\romannumeral #1@}
\newcommand{\mv}[1]{\mbox{\boldmath{$ #1 $}}}

\makeatother

\newtheorem{definition}{Definition}

\begin{document}
\title{Movable-Antenna Position Optimization for Physical-Layer Security via Discrete Sampling}
\author{\IEEEauthorblockN{Weidong Mei\IEEEauthorrefmark{1}, Xin Wei\IEEEauthorrefmark{1}, Yijie Liu\IEEEauthorrefmark{2}, Boyu Ning\IEEEauthorrefmark{1}, and Zhi Chen\IEEEauthorrefmark{1}}
\IEEEauthorblockA{\IEEEauthorrefmark{1}National Key Laboratory of Wireless Communications,}
\IEEEauthorblockA{\IEEEauthorrefmark{2}School of Information and Communication Engineering,\\ University of Electronic Science and Technology of China, Chengdu, China}
Emails: wmei@uestc.edu.cn; xinwei@std.uestc.edu.cn; 2022010909010@std.uestc.edu.cn; \\boydning@outlook.com; chenzhi@uestc.edu.cn}
\maketitle

\begin{abstract}
Fluid antennas (FAs) and mobile antennas (MAs) are innovative technologies in wireless communications that are able to proactively improve channel conditions by dynamically adjusting the transmit/receive antenna positions within a given spatial region. In this paper, we investigate an MA-enhanced multiple-input single-output (MISO) secure communication system, aiming to maximize the secrecy rate by jointly optimizing the positions of multiple MAs. Instead of continuously searching for the optimal MA positions as in prior works, we propose to discretize the transmit region into multiple sampling points, thereby converting the continuous antenna position optimization into a discrete sampling point selection problem. However, this point selection problem is combinatory and thus difficult to be optimally solved. To tackle this challenge, we ingeniously transform this combinatory problem into a recursive path selection problem in graph theory and propose a partial enumeration algorithm to obtain its optimal solution without the need for high-complexity exhaustive search. To further reduce the complexity, a linear-time sequential update algorithm is also proposed to obtain a high-quality suboptimal solution. Numerical results show that our proposed algorithms yield much higher secrecy rates as compared to the conventional FPA and other baseline schemes.
\end{abstract}

\section{Introduction}
The inherent broadcasting nature of wireless channels renders wireless communication systems particularly susceptible to eavesdropping attacks. To address this issue, secure multi-antenna beamforming has been developed to bolster the security of wireless communications by increasing the difference in signal strength between legitimate users and eavesdroppers\cite{mukherjee2014principles,liu2016physical,chen2016survey,wu2018survey}. However, existing multi-antenna beamforming solutions are limited to adapting to wireless channels without the capability to actively reshape them, which thus may result in suboptimal secure communication performance, especially in scenarios where the channels of legitimate users and eavesdroppers are highly correlated.

To overcome this limitation, fluid antenna (FA) and movable antenna (MA) technologies have emerged as promising solutions owing to their ability to dynamically adjust the positions of transmit/receive antennas within a given region\cite{wong2020fluid,wong2023fluid,zhu2024movable,zhu2024modeling}. As compared to the conventional fixed-position antennas (FPAs), FAs/MAs can proactively reshape the wireless channels into a more favorable condition for secure communications by circumventing the positions that may experience deep fading for legitimate users and/or substantial information leakage to eavesdroppers. Inspired by the promising benefits of the FA/MA technology, prior studies have delved into the antenna position optimization problem in FA-/MA-assisted secure communications \cite{hu2024secure,cheng2024enabling,tang2023fluid,tang2024secure,ding2024secure}. Specifically, in \cite{hu2024secure} and \cite{cheng2024enabling}, the authors aimed to optimize the antenna positions to maximize the secrecy rate in a multiple-input single-output (MISO) communication system with a single eavesdropper, based on a field-response channel model in the angular domain. While the authors in \cite{tang2023fluid,tang2024secure,ding2024secure} further introduced artificial noise in the transmission to degrade the received signal quality at the eavesdropper and solved the associated antenna position optimization problems. 

\begin{figure}[!t]
\centering
\includegraphics[width=2.7in]{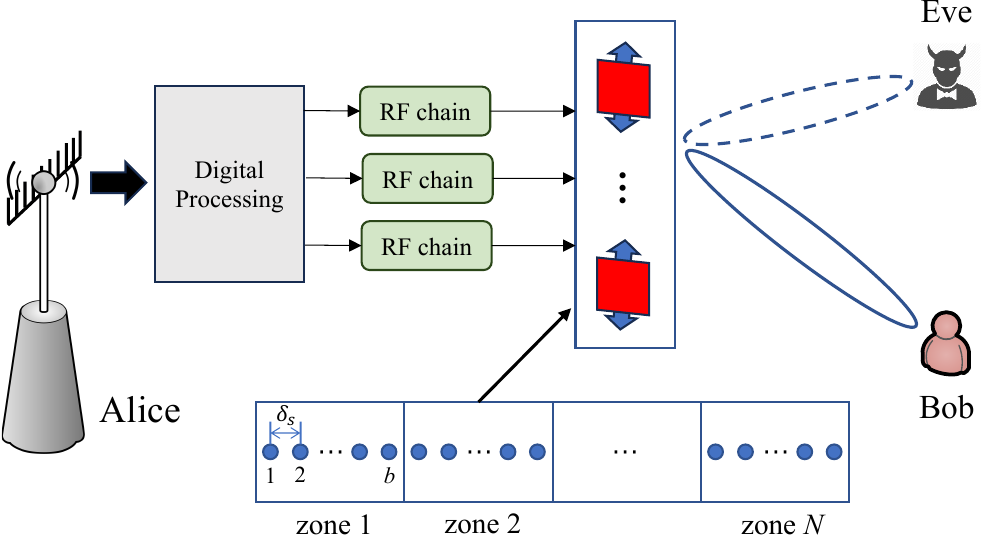}
\DeclareGraphicsExtensions.
\caption{MA-enhanced secure communication system with discrete sampling.}\label{sysmod}
\vspace{-9pt}
\end{figure}
However, all of the above works focused on searching for the optimal antenna positions in a continuous space, which may be difficult due to the highly nonlinear channel responses with respect to (w.r.t.) the antenna positions. Alternatively, another approach is to discretize the transmit or receive region into multiple ports/sampling points, over which FAs/MAs can be moved. Based on the channel state information (CSI) for each port or sampling point, optimal antenna positions can be determined by solving a port/point selection problem\cite{chai2022port} while ensuring minimum antenna spacing to prevent mutual coupling. As compared to the continuous antenna position optimization, employing a discrete port/point selection method is generally more straightforward to implement. However, it may entail much higher computational complexity in finding optimal solutions due to the typically high level of discretization required to reduce the performance loss compared to continuous position optimization. In \cite{mei2024movable} and \cite{wu2023movable}, the authors proposed a graph-based algorithm and a generalized Bender's decomposition method to solve the point selection problem optimally for single-user and multi-user MISO communication systems, respectively. However, to the best of our knowledge, there is no existing work studying the port/point selection problem for a secure communication system.

To fill in this gap, we study the sampling point selection problem for an MA-enhanced MISO secure communication system in this paper, with multiple MAs equipped in a confined region at the transmitter. To facilitate the antenna movement, we also partition the transmit region into several zones, each designated for the movement of a specific MA, as shown in Fig.\,\ref{sysmod}. By discretizing the transmit region into a multitude of discrete sampling points, our goal is to maximize the secrecy rate by jointly optimizing the transmit beamforming and the position of each MA over the sampling points within its designate zone. We first derive the optimal transmit beamforming in closed-form for a given set of sampling points. However, the remaining point selection optimization problem is combinatory and thus difficult to be optimally solved. To tackle this difficulty, we ingeniously model the discrete points as vertices in a graph, thereby recasting the point selection problem as an equivalent recursive path selection problem for a multipartite graph. However, achieving optimal path selection still requires exponential complexity w.r.t. the number of MAs. To mitigate the high enumeration complexity, we propose a solution bounding method allowing for more efficient partial enumeration by properly discarding some solution sets that cannot yield the optimal solution. To further reduce the computational complexity, a linear-time suboptimal sequential update algorithm is also proposed by sequentially selecting the discrete points for MAs. Numerical results demonstrate that our proposed algorithms significantly outperform the conventional FPAs and other heuristic schemes.

{\it Notations:} Bold symbols in capital letter and small letter denote matrices and vectors, respectively. The conjugate transpose of a vector or matrix is denoted as ${(\cdot)}^{H}$. ${\mathbb{R}}^n$ (${\mathbb{C}}^n$) denotes the set of real (complex) vectors of length $n$. $\lVert \mv a \rVert$ denotes the Euclidean norm of the vector $\mv a$. For a complex number $s$, $s \sim \mathcal{CN}(0,\sigma^2)$ means that it is a circularly symmetric complex Gaussian (CSCG) random variable with zero mean and variance $\sigma^2$. For two sets $A$ and $B$, $A \cup B$ and $A \cap B$ denote their union and intersection, respectively. $\lvert A \rvert$ denotes the cardinality of the set $A$. ${n \choose k} = \frac{n!}{k!(n-k)!}$ denotes the number of ways to choose $k$ elements from a set of $n$ elements. ${\cal O}(\cdot)$ denotes the order of complexity. ${\bf{1}}$ denotes an all-one vector.

\begingroup
\allowdisplaybreaks
\section{System Model and Problem Formulation}
\subsection{System Model}
As shown in Fig.\,\ref{sysmod}, we consider a MISO secure communication system, where a transmitter communicates with a legitimate receiver in the presence of an unauthorized receiver/eavesdropper. For convenience, we refer to the transmitter, the legitimate receiver, and the eavesdropper as {\it Alice}, {\it Bob}, and {\it Eve}, respectively. We assume that Alice is equipped with $N$ MAs, while Bob and Eve are equipped with a single FPA. We consider a linear transmit MA array of length $L$ at Alice, over which the positions of its $N$ MAs can be flexibly adjusted. Let ${\cal N}=\{1,2,\cdots,N\}$ denote the set of all MAs. We assume that the channel from Alice to Bob/Eve is slow-varying, such that the MAs in ${\cal N}$ can be moved to their respective optimized positions with negligible time as compared to the channel coherence time.

For the ease of implementing antenna movement, we uniformly sample the transmit array into $M\;(M \gg N)$ discrete positions, with an equal spacing between any two adjacent sampling points given by $\delta_s=L/M$, as shown in Fig.\,\ref{sysmod}. Hence, the position of the $m$-th sampling point is given by $s_m=\frac{mL}{M}, m \in {\cal M}=\{1,2,\cdots,M\}$, with ${\cal M}$ denoting the set of all sampling points within the MA array. As such, the position of each MA can be selected as one of the sampling points in ${\cal M}$. Let $a_n, a_n \in \cal M$ denote the index of the selected sampling point for the $n$-th MA. Thus, the position of the $n$-th MA can be expressed as $s_{a_n}=\frac{a_nL}{M}, n \in \cal N$. Furthermore, as shown in Fig.\,\ref{sysmod}, to enable fast antenna movement and also facilitate the antenna position optimization, we uniformly divide the transmit MA array into $N$ zones, such that the $n$-th MA can only be moved within the $n$-th zone\cite{chen2023joint}. Let $b=M/N$ denote the number of sampling points in each zone, which is assumed to be an integer for simplicity. Hence, it must hold that
\begin{equation}\label{zone}
b(n-1)+1 \le a_n \le bn, n \in {\cal N}.	
\end{equation}
To avoid the mutual coupling between MAs with a finite size, we consider a minimum distance, $d_{\min}$, between any pair of MAs. Thus, it should also hold that
\begin{equation}\label{dmin}
a_j - a_i \ge a_{\min}, \forall i,j \in {\cal N}, j > i,
\end{equation} 
where $a_{\min}\!=\! d_{\min}/\delta_s \gg 1$, which is assumed to be an integer. It follows that the MA position optimization is equivalent to the sampling point selection from $\cal M$ subject to (\ref{dmin}). 

Denote by $h_{m,B} \in {\mathbb{C}}$ and $h_{m,E} \in {\mathbb{C}}, m \in \cal M$ the baseband-equivalent channels from the $m$-th sampling point to Bob and Eve, respectively. To investigate the performance limit of our proposed algorithms, we assume that $h_{m,B}$'s and $h_{m,B}$'s are known at Alice, while in practice they can be acquired by applying various channel estimation techniques dedicated to MAs\cite{ma2023compressed,xu2024channel}. Based on the above, the channels from Alice to Bob and Eve are respectively expressed as 
\begin{align}
	{\mv h}_B(\{a_n\})=[h_{a_1,B},h_{a_2,B},\cdots,h_{a_N,B}]^H,\nonumber\\
	{\mv h}_E(\{a_n\})=[h_{a_1,E},h_{a_2,E},\cdots,h_{a_N,E}]^H.\nonumber
\end{align}

Let ${\mv w}_t \in {\mathbb C}^{N \times 1}$ and $P_t$ denote the transmit beamforming vector and transmit power at Alice, respectively, with $\lVert {\mv w}_t \rVert^2 \le P_t$. As such, the achievable secrecy rate is
\begin{align}
R_s({\mv w}_t,\{a_n\}) = \log_2\left(1+\frac{\lvert {\mv w}^H_t{\mv h}_B(\{a_n\}) \rvert^2}{\sigma^2}\right) - \nonumber\\
\log_2\left(1+\frac{\lvert {\mv w}^H_t{\mv h}_E(\{a_n\})\rvert^2}{\sigma^2}\right),\label{secrate}
\end{align}
where $\sigma^2$ denotes the receiver noise power at Bob/Eve.\vspace{-3pt}

\subsection{Problem Formulation}
In this paper, we aim to maximize the secrecy rate in (\ref{secrate}) by jointly optimizing the transmit beamforming and the MA positions over the discrete sampling points, i.e., ${\mv w}_t$ and $\{a_n\}_{n \in \cal N}$. The corresponding optimization problem is formulated as
\begin{align}
{\text{(P1)}}\; \mathop {\max}\limits_{{\mv w}_t,\{a_n\}}&\; R_s({\mv w}_t,\{a_n\}) \nonumber\\
\text{s.t.}\;\;&b(n-1)+1 \le a_n \le bn, n \in {\cal N}, \label{op2a}\\
&a_j - a_i \ge a_{\min}, \forall i,j \in {\cal N}, j>i, \label{op2b}\\
&\lVert {\mv w}_t \rVert^2 \le P_t.
\end{align}
It is worth noting that as compared to the continuous MA position optimization problems for secrecy rate maximization under the field-response based channel model (see e.g., \cite{hu2024secure,cheng2024enabling,tang2023fluid,tang2024secure}), (P1) is a discrete optimization problem that avoids the highly nonlinear channel expressions w.r.t. the MA positions. 

Note that for any given MA positions $\{a_n\}$, it can be shown that the optimal transmit beamforming for (P1) can be obtained by solving the following optimization problem, i.e.,
\begin{equation}\label{optBF}
	{\mv w}_t(\{a_n\})=\arg \mathop {\max}\limits_{\lVert {\mv w}_t \rVert^2 \le P_t} \log_2\left(\frac{\sigma^2+\lvert{\mv w}^H_t{\mv h}_B(\{a_n\})\rvert^2}{\sigma^2+\lvert{\mv w}^H_t{\mv h}_E(\{a_n\})\rvert^2}\right).
\end{equation}
Problem (\ref{optBF}) is equivalent to a generalized Rayleigh quotient problem, for which the optimal solution can be derived in closed form as
\begin{equation}\label{quotient}
{\mv w}_t(\{a_n\})=\sqrt{P_t}{\mv f}({\mv A}_B(\{a_n\}),{\mv A}_E(\{a_n\})),
\end{equation}
where ${\mv A}_B(\{a_n\})\!\!=\!\!\sigma^2{\mv I}_{N_t}\!+\!{\mv h}_B(\{a_n\}){\mv h}^H_B(\{a_n\})$, ${\mv A}_E(\{a_n\})\!\!=\!\!\sigma^2{\mv I}_{N_t}+{\mv h}_E(\{a_n\}){\mv h}^H_E(\{a_n\})$, and ${\mv f}({\mv A}_B(\{a_n\}),{\mv A}_E(\{a_n\}))$ denotes the eigenvector corresponding to the maximum eigenvalue of the matrix ${\mv A}^{-1}_E(\{a_n\}){\mv A}_B(\{a_n\})$. 

By substituting (\ref{quotient}) into (\ref{secrate}), the maximum secrecy rate for any given MA positions $\{a_n\}$ can be obtained as
\begin{equation}\label{MaxRate}
R_s(\{a_n\}) = \log_2\lambda_{\max}({\mv A}_B(\{a_n\}),{\mv A}_E(\{a_n\})),
\end{equation}
where $\lambda_{\max}({\mv X},{\mv Y})$ denotes the maximum eigenvalue of the matrix ${\mv Y}^{-1}{\mv X}$. Then, (P1) can be simplified as
\begin{align}
{\text{(P2)}}\; \mathop {\max}\limits_{\{a_n\}}&\; \log_2\lambda_{\max}({\mv A}_B(\{a_n\}),{\mv A}_E(\{a_n\})) \nonumber\\
\text{s.t.}\;\;& {\text{(\ref{op2a})-(\ref{op2b})}}.
\end{align}

However, (P2) is still challenging to be optimally solved due to the intricate relationship between its objective function and the MA positions $\{a_n\}$, as well as its combinatorial nature. One straightforward approach to optimally solve (P2) is by enumerating all possible MA positions. However, this may incur exorbitant complexity (e.g., in the order of ${M-(N-1)a_{\min} \choose N}$ if $a_{\min} \ge b$), which may not be applicable to a large size of antenna array with large $M$ and/or $N$ values in practice. 

\section{Proposed Algorithms for (P2)}\label{alg}
To solve (P2), we propose in this section a more efficient graph-based optimal solution based on partial enumeration and a lower-complexity suboptimal solution, respectively.

\subsection{Optimal Solution by Partial Enumeration}
First, we propose a recursive algorithm to efficiently enumerate all feasible solutions to (P2) based on a graph-based formulation, jointly with a bounding approach to properly discard some undesired solutions in the enumeration, as specified below.

{\it 1) Graph-based Formulation:} Specifically, we construct a directed weighted graph $G=(V,E)$. The vertex set $V$ is given by the set of all sampling points, i.e., $V={\cal M}$. Without loss of optimality, we consider that the MA indices are selected in order from one sampling point $i$ in a zone to a farther point $j$ from the reference position 0 in an adjacent zone. Accordingly, we add an edge from vertex $i$ to vertex $j$ if and only if $j-i \ge a_{\min}$ and they are located in two adjacent zones, corresponding to the constraints (\ref{op2a}) and (\ref{op2b}). Note that for any vertex/sampling point $i, i \in \cal M$, the index of its located zone is given by $z_i = \lceil i/b \rceil$. Hence, the edge set $E$ is defined as
\begin{equation}\label{edgeset}
E=\{(i,j)|j-i \ge a_{\min}, z_j=z_i+1, i, j \in {\cal M}\}.
\end{equation}
It can be verified that 
\begin{equation}
\lvert E \rvert=
\begin{cases}
	(N-1)\frac{(2b-a_{\min})(2b-a_{\min}+1)}{2},\!&{\text{if}}\;a_{\min} \ge b\\
	(N-1)\left(b^2-\frac{1}{2}a_{\min}(a_{\min}-1)\right),\!&{\text{otherwise}}
\end{cases}.
\end{equation}

To relate the graph $G$ to (P2), we introduce the following definition.
\begin{definition}
A $K$-partite graph refers to a graph whose vertices can be partitioned into $K$ disjoint sets, such that there is no edge between any two vertices within the same set.	
\end{definition}

It follows from Definition 1 that $G$ must be an $N$-partite graph with its $n$-th disjoint set given by $V_n=\{m|b(n-1)+1 \le m \le bn\}$. As such, each feasible solution to (P2) corresponds to an $N$-vertex path from one vertex in $V_1$ to another vertex in $V_N$. 

{\it 2) Feasible Solution Enumeration:} Based on the unique $N$-partite property of $G$, we next propose a recursive method to enumerate all feasible solutions to (P2). To this end, we show that any desired $N$-vertex path in $G$ can be recursively constructed based on the paths with a smaller number of vertices. Note that its $N$ vertices must be selected from the $N$ disjoint sets $V_n, n \in \cal N$, respectively. Without loss of optimality, we assume that its $n$-th vertex is selected from $V_n$. Accordingly, let $\Omega_r, r \le N$ denote the set of all $r$-vertex paths from one vertex in $V_1$ to another vertex in $V_r$ in $G$, with the $s$-th vertex of each path in $\Omega_r$ selected from $V_s, s=1,2,\cdots,r$. Obviously, we have $\Omega_1=V_1$. Moreover, for each path in $\Omega_r, r \le N$, if there exists a vertex in $V_{r+1}$ which is adjacent to its $r$-th vertex, then a new path in $\Omega_{r+1}$ can be constructed by appending the vertex to this path. As such, based on the initial condition for $\Omega_1$ and the recursion for $\Omega_r, r \le N$, all $N$-vertex paths in $G$ can be enumerated in the set $\Omega_N$. 

{\it 3) Feasible Solution Bounding:} Although the above recursive method can enumerate all feasible solutions to (P2), its worst-case complexity can be prohibitive for large $M$ and/or $N$ values. Next, we propose a bounding method to determine an upper performance bound by any path in $\Omega_r, r \le N$. Evidently, if such an upper bound is no larger than the performance by any incumbent feasible solution to (P2) (e.g., the suboptimal solution proposed in the next subsection), then the recursion from this path can be terminated to reduce the enumeration complexity. Let $\Gamma_r=({\tilde a}_1,{\tilde a}_2,\cdots,{\tilde a}_r)$ denote any path in $\Omega_r, 1 \le r \le N$, with ${\tilde a}_s \in V_s, 1 \le s \le r$. Then, the maximum secrecy rate achievable by $\Gamma_r$ can be obtained by replacing $a_s$ in (P2) with ${\tilde a}_s \in V_s, 1 \le s \le r$, respectively. To determine an upper bound on this maximum secrecy rate, note that it must be no larger than the maximum achievable rate at Bob under the maximum-ratio transmission (MRT), i.e., ${\mv w}_t = \sqrt{P_t}{\mv h}_B(\{a_n\})/\lVert {\mv h}_B(\{a_n\}) \rVert$ with $a_s={\tilde a}_s, 1 \le s \le r$. Hence, the desired upper bound can be determined by solving the following optimization problem, i.e.,
\begin{align}
\mathop {\max}\limits_{\{a_n\}_{r< n \le N}}&\; \log_2\left(\frac{P_t}{\sigma^2}\left(\sum\limits_{n=1}^r{\lvert h_{{\tilde a}_n}\rvert^2}+\sum\limits_{n=r+1}^N{\lvert h_{a_n} \rvert^2}\right)\right)\nonumber\\
\text{s.t.}\;\;&b(n-1)+1 \le a_n \le bn, r< n \le N,\nonumber\\
&a_{r+1} - {\tilde a}_r \ge a_{\min},\nonumber\\
&a_j - a_i \ge a_{\min}, r< i<j \le N,\label{SPP}
\end{align}
where the term $\sum\nolimits_{n=1}^r{\lvert h_{{\tilde a}_n}\rvert^2}$ in the objective function can be viewed as a constant. Note that the upper bound obtained by solving (\ref{SPP}) becomes tight if the MRT based on the Alice-to-Bob channel, ${\mv h}_B(\{a_n\})$, can null the received signal at Eve, which may be easier to be achieved for MAs than FPAs\cite{zhu2024movable}.

Problem (\ref{SPP}) can be optimally solved in polynomial time by applying a similar algorithm as in our previous work\cite{mei2024movable}. Thus, we only outline the main steps of solving it. Specifically, we add a ``dummy'' vertex $M+1$ to $G$ and add an edge from each vertex in $V_N$ to vertex $M+1$. As such, the vertex set becomes $\tilde V = V \cup \{M+1\}$, while the edge set $E$ becomes
\begin{equation}
\tilde E=E \cup \{(j,M+1)|j \in V_N\}.
\end{equation}
Moreover, for the new graph $\tilde G=(\tilde V,\tilde E)$, we set the weights of its edge $(i,j), (i,j) \in \tilde E$ as $\tilde W_{i,j}=-\lvert h_i \rvert^2$. By dropping the logarithm and other irrelevant constant scalars, the objective value of problem (\ref{SPP}) by any MA indices $a_n, r<n\le N$, i.e., $\sum\nolimits_{n=r+1}^N{\lvert h_{a_n} \rvert^2}$, is equal to the negative sum of edge weights of the path ${\tilde a}_r \rightarrow a_{r+1}\rightarrow a_{r+2}\rightarrow \cdots \rightarrow a_N \rightarrow M+1$. Hence, problem (\ref{SPP}) is equivalent to finding the shortest path (i.e., with the minimum sum of the weights of the constituent edges) from vertex ${\tilde a}_r$ to vertex $M+1$ in $\tilde G$, for which some celebrated shortest-path algorithms of polynomial complexity can be applied, e.g., Dijkstra algorithm.

Denote by ${\bar a}_n, n \in \cal N$ any given feasible solution to (P2), with its achieved secrecy rate given by ${\bar R}_s=R_s(\{{\bar a}_n\})$ based on (\ref{MaxRate}). Let $T(\Gamma_r)$ denote the optimal value of (\ref{SPP}) by the path $\Gamma_r$. It is evident that if $T(\Gamma_r) \le {\bar R}_s$, then there will be no need to execute any further recursion from $\Gamma_r$. Otherwise, the recursion proceeds to the $(r+1)$-vertex path, until all $N$-vertex paths are enumerated with the above bounding method. It is worth noting that to maximally reduce the enumeration, ${\bar R}_s$ is expected to be sufficiently large. In this paper, we apply the suboptimal solution presented in Section \ref{sub} to determine ${\bar R}_s$, which is shown able to achieve near-optimal performance in Section \ref{sim} based on simulation. Furthermore, each time an $N$-vertex path is found, e.g., $\Gamma_N$, we compare its achieved secrecy rate, i.e., $R_s(\Gamma_N)$, with ${\bar R}_s$. If the former is larger, we can update $\{{\bar a}_n\}$ and ${\bar R}_s$ as $\Gamma_N$ and $R_s(\Gamma_N)$, respectively, to further reduce the enumeration complexity. 

Finally, we output $\{{\bar a}_n\}$ and ${\bar R}_s$ as the optimal solution to (P2) and its optimal value, respectively, after the enumeration is completed. The main procedures of our proposed algorithm for solving (P2) are summarized in Algorithm \ref{Alg1}, where a function ``$\textsc{RecEnum}$'' is defined and recursively called to achieve the recursive enumeration. \vspace{-3pt}
\begin{algorithm}
  \caption{Proposed Partial Enumeration Method for Solving (P2)}\label{Alg1}
  \begin{algorithmic}[1]
  \State Determine a feasible solution to (P2), $\{{\bar a}_n\}$, and calculate its achieved secrecy rate, ${\bar R}_s=R_s(\{{\bar a}_n\})$, based on (\ref{MaxRate}).
  \State Initiate $r=1$, $\Gamma_0=\emptyset$, and $\tilde\Omega_N=\emptyset$.
  \State Execute $\textsc{RecEnum\,}(r,\Gamma_0)$.
  \State Output $\{{\bar a}_n\}$ and ${\bar R}_s$.
  \Function{RecEnum\,}{$r,\Omega$}
    \If {$r=N+1$}
    \State Add $\Gamma_N$ to the set $\tilde\Omega_N$.
    \If {$R_s(\Gamma_r) > {\bar R}_s$}
    \State Update $\{{\bar a}_n\}=\Gamma_N$ and ${\bar R}_s=R_s(\Gamma_N)$.
    \EndIf
    \Else
    \State Initialize $s=1$.
    \While {$s \le N$}
    \State \begin{varwidth}[t]{0.81\linewidth}Append the $s$-th vertex in $V_r$ to $\Gamma_{r-1}$ and obtain a $r$-vertex path, $\Gamma_r$.\end{varwidth}\vspace{3pt}
    \State \begin{varwidth}[t]{0.81\linewidth}Calculate $T(\Gamma_r)$ by solving problem (\ref{SPP}) with the Dijkstra algorithm.\end{varwidth}\vspace{3pt}
    \If {$T(\Gamma_r) > R_s(\{{\bar a}_n\})$}
    \State Execute $\textsc{RecEnum\,}(r+1,\Gamma_r)$.
    \EndIf
    \State Update $s=s+1$.
    \EndWhile
    \EndIf
  \EndFunction
  \end{algorithmic}
\end{algorithm}\vspace{-9pt}

\subsection{Suboptimal Solution by Sequential Update}\label{sub}
Although the optimal solution to (P2) can be obtained by Algorithm \ref{Alg1}, its worst-case complexity, albeit rarely encountered, is equal to the complexity of full enumeration. To address this issue, we propose a suboptimal sequential update algorithm to solve (P1) with lower complexity, by sequentially selecting the sampling points for MAs.

Specifically, let $\{a'_n, n \in {\cal N}\}$ denote a set of initial MA positions, with $a'_n \in V_n$. Consider that the $n$-th MA position, i.e., $a'_n$, needs to be updated in the $n$-th iteration of the sequential search. Then, the following optimization problem should be solved,
\begin{equation}\label{an}
\mathop {\max}\limits_{a_n \in \Psi_n}\!\log_2\lambda_{\max}({\mv A}_B(a_n;\{a'_i\}_{i \ne n}),{\mv A}_E(a_n;\{a'_i\}_{i \ne n})),
\end{equation}
where 
\begin{equation}\label{psi}
	\Psi_n\!=\!\{m|m \in V_n, m-a'_{n-1}\!\ge\!a_{\min}, a'_{n+1}-m\!\ge\!a_{\min}\},
\end{equation}
for $2 \le n \le N-1$, and we set $\Psi_1=\{m|m \in V_1, a'_2-m \ge a_{\min}\}$ and $\Psi_N=\{m|m \in V_N, m-a'_{N-1} \ge a_{\min}\}$. 

Let $a_n^*$ denote the optimal solution to problem (\ref{an}), which can be determined by performing an enumeration over $\Psi_n$. Next, we update $a_n'=a_n^*$ and the update of the $(n+1)$-th sampling point follows. Since $\lvert \Psi_n \rvert \le b$, the complexity of our proposed sequential update algorithm is in the order ${\cal O}(M)$, which is thus much lower than that of the partial enumeration algorithm in general. However, it should be mentioned that the sequential update algorithm may only achieve suboptimal performance, as the sets $\Psi_n, n \in \cal N$ depend on the initial sampling point selection and the order of the selected sampling points, and thus some optimal sampling points may be eliminated in the update of these sets. The main steps of the sequential update algorithm are summarized in Algorithm \ref{Alg}. 
\begin{algorithm}
  \caption{Sequential Update Algorithm for Solving (P2)}\label{Alg}
  \begin{algorithmic}[1]
    \State Initialize $n=1$, $a'_n, n \in {\cal N}$, and $\Psi_1$.
    \While {$n \le N$}
    \State Determine $a_n^*$ by solving (\ref{an}) and update $a'_n=a_n^*$.
    \State Determine $\Psi_{n+1}$ based on (\ref{psi}).
    \State Update $n=n+1$.
    \EndWhile
    \State Output $\{a'_n\}$ as the optimized MA positions.
    \end{algorithmic}
\end{algorithm}
\vspace{-3pt}
\section{Numerical Results}\label{sim}
In this section, we provide numerical results to evaluate the performance of our proposed optimal graph-based algorithm and the suboptimal sequential update algorithm. Unless otherwise stated, the simulation settings are as follows. The carrier frequency is 5 GHz, and thus the wavelength is $\lambda=0.06$ meter (m). The number of transmit MAs is $N = 6$, while the length of the linear transmit array is $L=0.36\;{\text{m}}=6\lambda$. The minimum distance between any two MAs is set to $d_{\min} = \lambda/2$. Let $D_B$ ($D_E$) and $\alpha$ denote the distance from Alice to Bob (Eve) and the path-loss exponent, respectively, which are set to $D_B=100$ m, $D_E=100$ m, and $\alpha=2.8$. To generate the channels for the sampling points, i.e., $h_m, m \in \cal M$, we consider the field-response channel model in \cite{hu2024secure,cheng2024enabling,tang2023fluid,tang2024secure}, with the number of transmit paths set to 9 for both Bob and Eve. Let $\gamma_i$ denote the channel response coefficient for the $i$-th transmit path, which is assumed to follow CSCG distribution with $\gamma_i \sim {\cal{CN}}(0,\beta D^{-\alpha}l_i)$, where $\beta$ denotes the path loss at the reference distance of 1 m, and $l_i$ denotes the ratio of the average power gain of the $i$-th transmit path to that of all transmit paths. We set $\beta=-46$ dB, while the values of $l_i$'s are randomly generated and then normalized to satisfy $\sum\nolimits_{i}l_i=1$. The angle of departure (AoD) from the transmit array to each transmit path (either for Bob or Eve) is assumed to be a uniformly distributed variable over $[0,\pi]$. The transmit signal-to-noise ratio (SNR) is $P_t/\sigma^2=100$ dB, with $\sigma^2$ denoting the receiver noise power. 

We compare our proposed algorithms with the following three benchmarks:
\begin{itemize}
\item {\bf{Benchmark 1: MRT without considering Eve.}} In this benchmark, the transmit beamforming and MA positions at Alice are optimized to maximize the received signal power at Bob only under the MRT.
\item {\bf{Benchmark 2: Channel difference maximization.}} In this benchmark, the MA positions are optimized to maximize ${\bf{1}}^T({\mv h}_B(\{a_n\})-{\mv h}_E(\{a_n\}))$, and then the transmit beamforming is determined based on (\ref{quotient}).
\item {\bf{Benchmark 3: FPA.}} In this benchmark, the position of the $n$-th antenna is fixed at the midpoint of the $n$-th zone, $n \in \cal N$.
\end{itemize}
Note that the MA positions in benchmark 1 and benchmark 2 can be determined by applying the optimal graph-based algorithm proposed in our prior work\cite{mei2024movable}. In the proposed sequential update algorithm, the sampling point selection is initialized as that by benchmark 3. All the results are averaged over 1000 independent channel realizations.

\begin{figure}[!t]
\centering
\centerline{\includegraphics[width=0.4\textwidth]{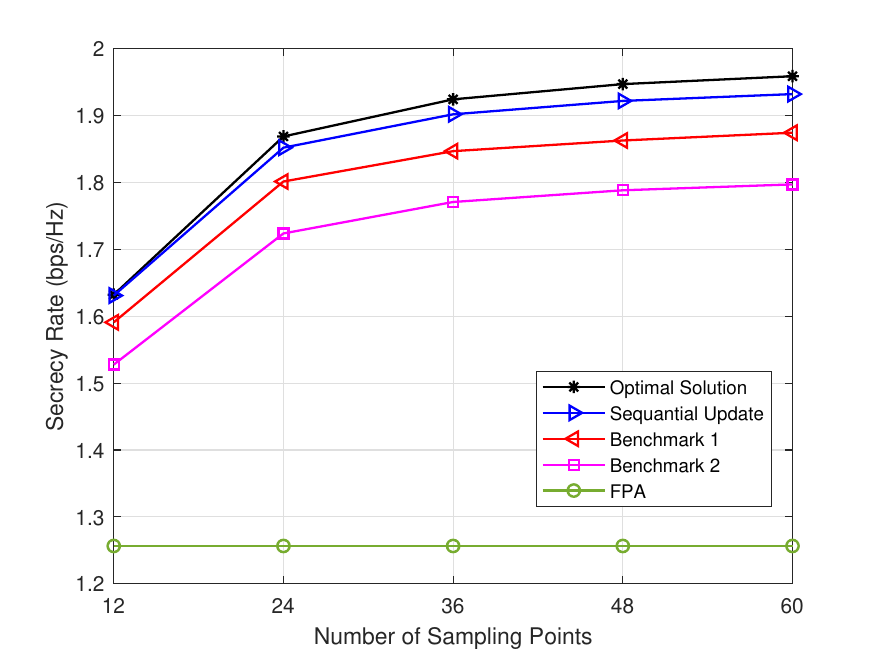}}
\caption{Secrecy rate versus the number of sampling points.}
\label{Fig_Samp}
\vspace{-8pt}
\end{figure}
First, we plot the secrecy rates by different algorithms versus the number of sampling points $M$ with $N = 6$ in Fig.\,\ref{Fig_Samp}. It is observed that the performance of the proposed algorithms and the first two benchmarks with MAs improves with $M$ thanks to the refined sampling resolution, resulting in an increasing performance gain over the FPAs. Particularly, our proposed algorithms can achieve considerably higher secrecy rates than the two benchmarks. However, when $M \ge 36$, further increasing $M$ cannot significantly enhance the secrecy rates, indicating that a sufficiently high resolution has been achieved. This also implies that a moderate number of sampling points is sufficient to achieve comparable performance to continuous searching. Moreover, the sequential update algorithm is observed to yield similar performance to the optimal solution, with a negligible performance gap 0.02 bps/Hz. 

\begin{figure}[!t]
\centering
\centerline{\includegraphics[width=0.4\textwidth]{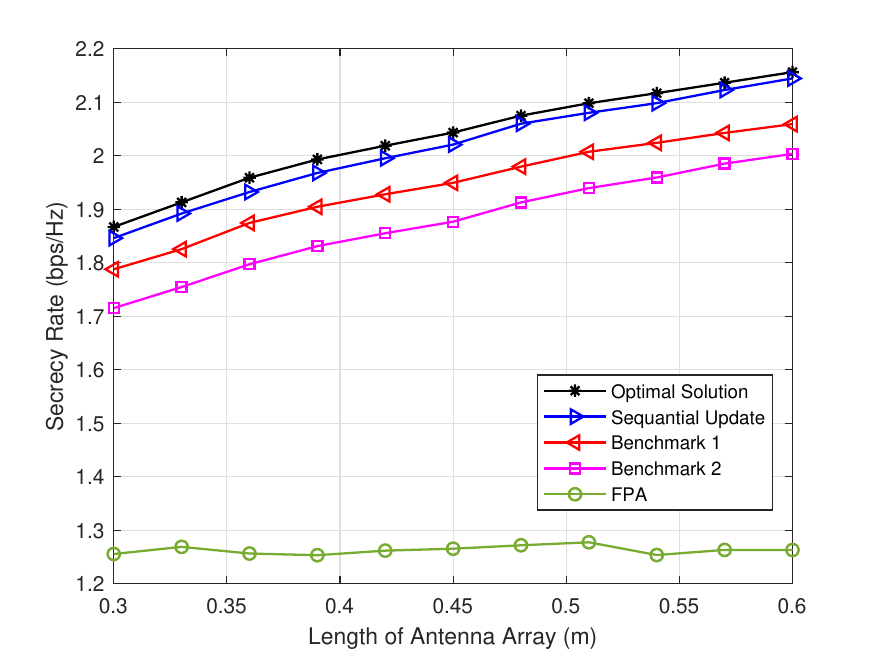}}
\caption{Secrecy rate versus the length of antenna arrays.}
\label{Fig_Length}
\vspace{-6pt}
\end{figure}
In Fig.\,\ref{Fig_Length}, we plot the secrecy rates by different algorithms versus the length of the transmit array, $L$, with $N = 6$ and the sampling resolution $\delta_s = 0.01$ m. Hence, there exist $M = L/\delta_s = 100L$ sampling points in total. It is observed that the performance of all schemes employing MAs increases with $L$, thanks to the enhanced degree of freedom for MA position optimization. In contrast, the secrecy rate by FPAs remains nearly constant as $L$ increases. This is attributed to the FPAs' inability to exploit the degree of freedom in antenna movement, which results in random channel conditions at their locations and consequently an approximately constant secrecy rate in average. Moreover, our proposed algorithms are observed to still yield much higher secrecy rates than the three benchmarks over all length of the transmit array considered. 

\begin{figure}[!t]
\centering
\centerline{\includegraphics[width=0.4\textwidth]{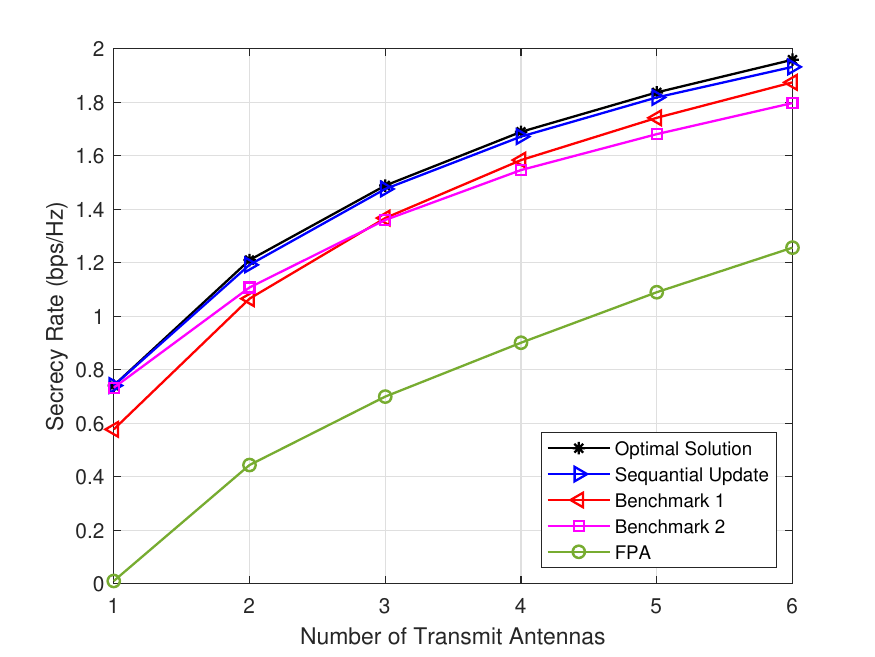}}
\caption{Secrecy rate versus the number of transmit MAs.}
\label{Fig_TxNums}
\vspace{-9pt}
\end{figure}
Finally, we plot in Fig.\,\ref{Fig_TxNums} the secrecy rates by different schemes versus the number of transmit antennas $N$, with $M=60$. It is observed that thanks to the improved beamforming gain, the performance of all schemes increases with $L$, and the maximum secrecy rate by the optimal algorithm increases from 0.74 bps/Hz to 1.96 bps/Hz, achieving a performance gain of 165\%. Moreover, our proposed algorithms are observed to significantly outperform the three benchmarks. However, it is observed that when $N=1$, benchmark 2 can achieve almost the same secrecy rate as the proposed algorithms. The possible reason is that in the absence of transmit beamforming in this single-antenna case, the optimal MA positions may depend more critically on the channel gain difference between Bob and Eve within the entire transmit array, which is aligned with the spirit of benchmark 2. This also results in the better performance of benchmark 2 compared to benchmark 1 when $N$ is small.

\section{Conclusion}
This paper considered an MA-enhanced secure communication system and optimized the MA positions over multiple designated zones in the transmit region to maximize the secrecy rate. By sampling the transmit region into multiple discrete points, we proposed a graph-based algorithm jointly with a solution bounding method to solve the point selection problem optimally by employing partial enumeration. Furthermore, a near-optimal sequential update algorithm was also proposed to solve the secrecy rate maximization problem in linear time. Numerical results show the superiority of our proposed algorithms to other baseline schemes given a moderate number of sampling points.\vspace{-6pt}

\bibliography{MA_new.bib}
\bibliographystyle{IEEEtran}
\end{document}